\documentclass[review]{elsarticle}

 \pdfoutput=1
 
\usepackage{hyperref}
\usepackage{lineno}
\usepackage{graphicx}
\usepackage{amsfonts}
\usepackage[utf8x]{inputenc} 
\usepackage{subfigure}
\modulolinenumbers[1]

\begin{document}

\begin{frontmatter}

\title{Multigap RPC time resolution \\ to \mbox{$511$ keV} annihilation photons}


\author[PV]{G. Belli$^{\dag}$}
\author[PV]{M. Gabusi}
\author[PV]{G. Musitelli$^{\dag}$}
\author[PV]{R. Nardò}
\author[PV]{\\S.P. Ratti}
	\cortext[cor1]{Corresponding author. Tel.: +39 0382 987637; fax: +39 0382 987752.}
\author[PV]{A.Tamborini\corref{cor1}}
	\ead{aurora.tamborini@unipv.it}
\author[PV]{P. Vitulo}

\address[PV]{University of Pavia-Department of Physics and INFN Section of Pavia, via Bassi $6$, $27100$ Pavia (Italy)}


\begin{abstract}

The time resolution of Multigap Resistive Plate Counters (MRPCs) to \mbox{$511$ keV} gamma rays has been investigated using a $^{22}$Na source and four detectors. The MRPCs time resolution has been derived from the Time-of-Flight information, measured from pairs of space correlated triggered events.  A GEANT$4$ simulation has been performed to analyze possible setup contributions and to support experimental results. 
A time resolution (FWHM) of $376$ ps and $312$ ps has been measured for a single MRPC with four $250$ $\mu$m gas gaps by considering respectively one and two independent pairs of detectors.\\
\end{abstract}

\begin{keyword}
Gaseous detectors; Multigap Resistive plate chambers (MRPC); Positron emission tomography (PET); Time of Flight (ToF). \\
\PACS 87.57.uk \sep 29.40.Cs
\end{keyword}

\end{frontmatter}


\section{Introduction} 

Resistive Plate Counters (RPCs) are gaseous detectors conceived in $1981$ \cite{Santonico} and used for charged particle detection in several experiments of high energy physics \cite{CMS,Cattani}. Multigap RPCs have been later introduced \cite{Zeballos} to improve the RPCs time resolution. Since then, an impressive amount of R$\&$D work  on these detectors has been carried out by several research groups and finally MRPCs have been employed or proposed in several experiments as Time of Flight detector \cite{Spegel, Bogomilov, Llope}. For such detectors, a time resolution down to $50$ ps has been measured for charged particles.\\
Soon the RPC based detector technology was also investigated for being exploited in gamma and neutron detection \cite{bib8, bib9} and MRPC detectors \cite{bib10, bib11} have been proposed for their use in Time of Flight Positron Emission Tomography \cite{bib12}.\\
Positron Emission Tomography (PET) is one of the most powerful and promising techniques of metabolic imaging in Nuclear Medicine: $2$D and $3$D images of tissues may be carried out by measuring the activity of a $\beta^{+}$ radio-tracer buildup into a biological volume \cite{bib13, bib14}.\\
In a PET scanner a ring of detectors collects the $511$ keV photons pairs produced by positron annihilation at rest: if the gamma pair is detected in coincidence, an ideal Line Of Response (LOR) can be reconstructed throughout the scanner.  Gathering the information of several LORs, a map of tracer concentration can be produced by means of reconstruction algorithms.\\
The Time of Flight (ToF) information of the two photons can be used to improve the reconstructed tomographic image \cite{bib15, bib16, bib17, bib18, bib19}. Currently, fast crystals such as LYSO, LSO, GSO and LaBr$_{3}$ are used in commercial Tof-PET scanners \cite{bib20}.\\
As concerning MRPCs, we already presented results on their gamma ToF capability by using two detectors \cite{bib21}. The following measurements refer to an improved setup with four detectors (section \ref{SubSec:1}).\\
The measurement setup will be described (section \ref{SubSec:4}) and experimental results will be shown (section \ref{Sec:5}) together with simulation outcomes (section \ref{Sec:6}).

\section{Materials and Methods}

\subsection{The detectors}\label{SubSec:1}

The used MRPCs detectors were made of a stack of $5$ glasses (\mbox{$200$ x $100$ mm$^{2}$} area) and $400$ $\mu$m thickness, separated by a nylon wire ($250$ $\mu$m diameter) that defines $4$ gaps. \\
Two thin resistive films ($180$ x $80$ mm$^{2}$ area, $70$ $\mu$m thickness) were applied on the external surface of the first and last glass, defining the high voltage and ground planes.  The glass stack was manually aligned, closed externally by two fiber glass plates (Printed Circuit Board, \mbox{$220$ x $100$ mm$^{2}$} area) and maintained in position using three Plexiglas bars kept under stress with springs.\\
Each stack was placed in a PMMA box (external dimension \mbox{$335$ x $195$ x $100$ mm$^{3}$}) and filled (at $15$ l/h) with a gas mixture composed by C$_{2}$H$_{2}$F$_{4}$, SF$_{6}$, iC$_{4}$H$_{10}$ ($85\%$, $10\%$, $5\%$).\\
Each MRPC module was designed to maintain high flexibility in terms of future modifications. As an example, two designs of the readout were produced and after a preliminary test using a single readout electrode with a sensitive area of $80$ x $140$ mm$^{2}$, the readout was partitioned in $16$ pads ($16$ x $16$ mm$^{2}$). Four detectors were finally instrumented for a total of $64$ channels.\\
Several efforts have been done for the electrodes' coating to allow for an optimal resistive film production, whose surface resistivity was of the order of \mbox{$20$ - $30$ M$\Omega$/$\Box$}. \\

\begin{figure}[htpb]
\centering
\includegraphics[scale=0.8]{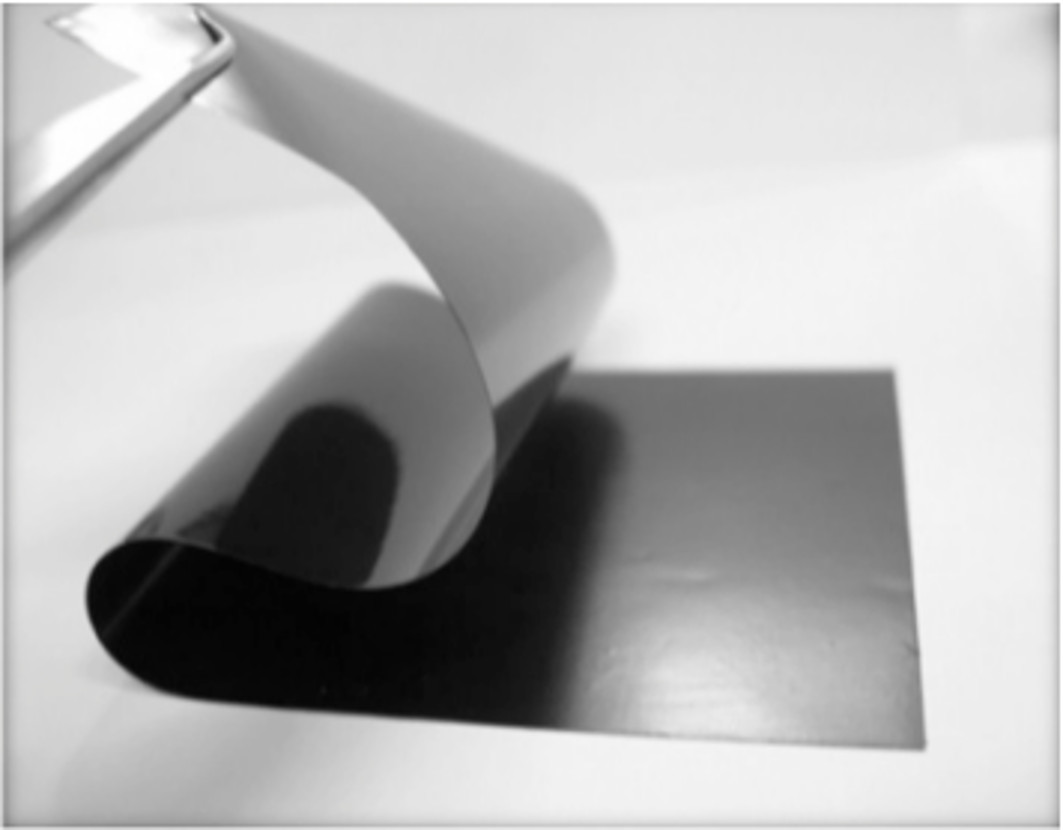}\\
\caption{Semi-resistive film for the high voltage application: $180$ x $80$ mm$^{2}$ area, $70$ $\mu$m thickness and surface resistivity around $20$-$30$ M$\Omega$/$\Box$.}
\label{fig:Resistive_Coating}
\end{figure}

The coating (Figure \ref{fig:Resistive_Coating}) was made of a polyester based paint\footnote{Supplied by PALINAL-Palini Vernici company (http://www.palinal.com)} enriched with Carbon Black (CB) nanoparticles. A study was carried out to find the right composition using different CB concentration \cite{bib22}.\\

\subsection{The electronics}\label{SubSec:2}
\begin{figure}[htpb]
\centering
\includegraphics[scale=1]{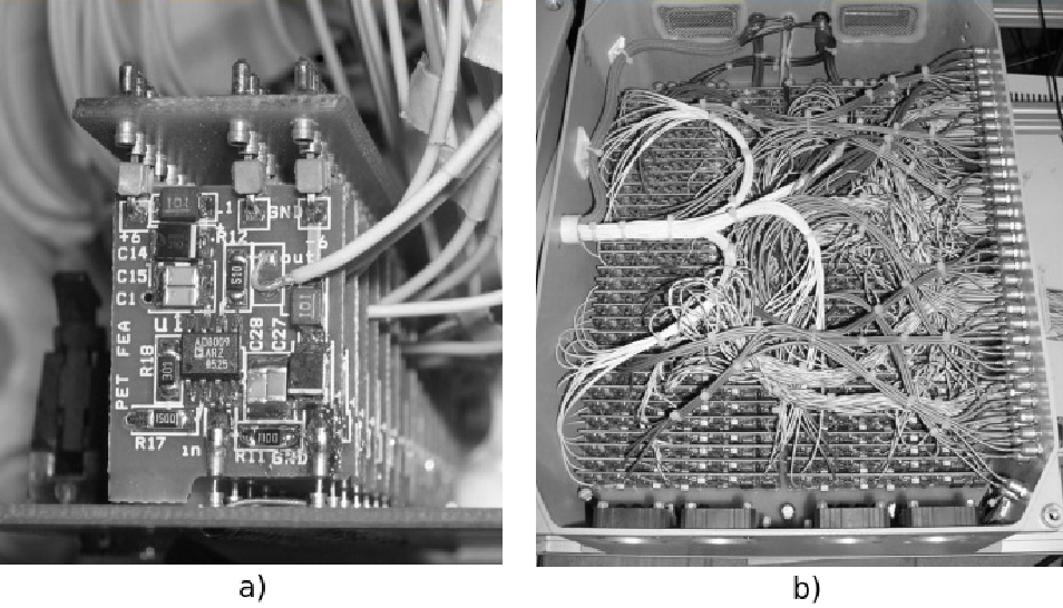}\\
\caption{Custom made electronics developed for the system: a) front-end stage; b) amplification and signal formation stage.}
\label{fig:Electronics}
\end{figure}

The signal from each pad of a MRPC was routed to the external boards through a flat cable.  A custom made front-end board (Figure \ref{fig:Electronics}a) provided initial signal amplification (x $2$). Signals from the front-end module fed a second amplification (x $3$) stage (Figure \ref{fig:Electronics}b). Different kind of outputs (analogical and digital ECL) could then be processed from the data acquisition chain (commercial NIM and VME modules).\\

\begin{figure}[htpb]
\centering
\includegraphics[scale=1]{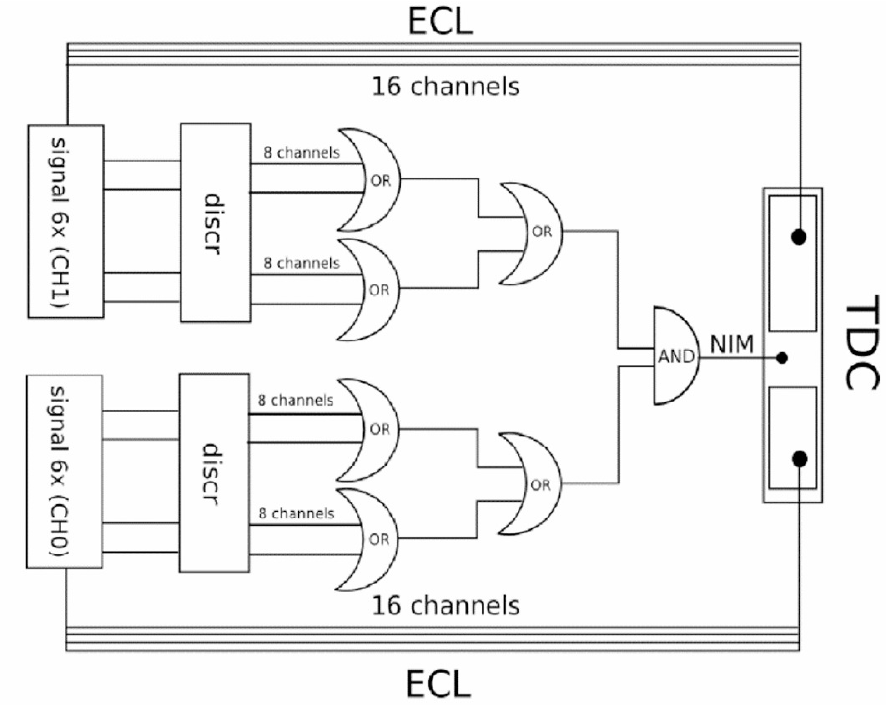}\\
\caption{Schematic representation of the electronics acquisition chain.}
\label{fig:Acquisition_Chain}
\end{figure}

While the analog outputs were used for triggering purposes, the ECL signals fed two VME CAEN V$1290$A TDCs that provided the inputs time stamps matching a valid trigger, as schematically shown in Figure \ref{fig:Acquisition_Chain} for two MRPC detectors. \\

\subsection{Channels synchronization}\label{SubSec:3}

The time stamp of a trigger pulse was provided by a coarse counter with a resolution of $25$ ns \cite{bib23} based on an on-board $40$ MHz oscillator. Thanks to an internal frequency multiplier (DLL) all the hits matching the coincidence window were recorded with a $25$ ps time resolution. \\

\begin{figure}[h]
\centering
\includegraphics[scale=1]{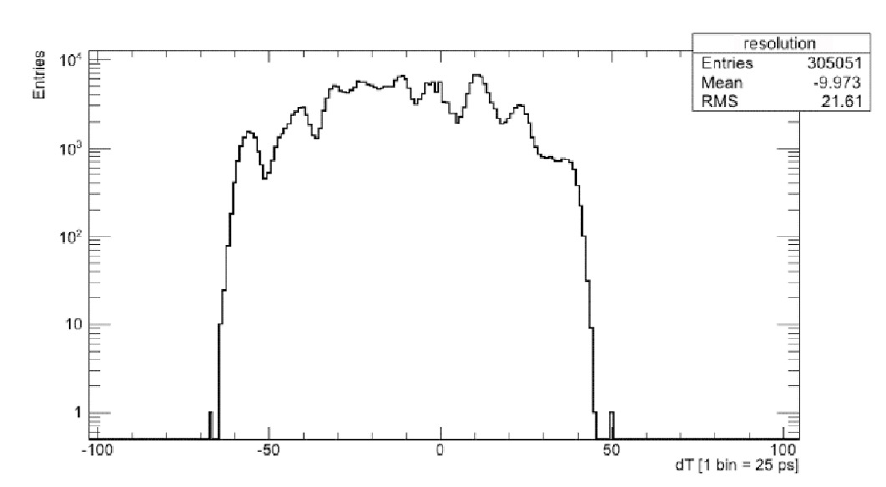}\\
\caption{Schematic representation of the electronics acquisition chain.}
\label{fig:Distribution_no_map}
\end{figure}

\begin{figure}[h]
\centering
\includegraphics[scale=1]{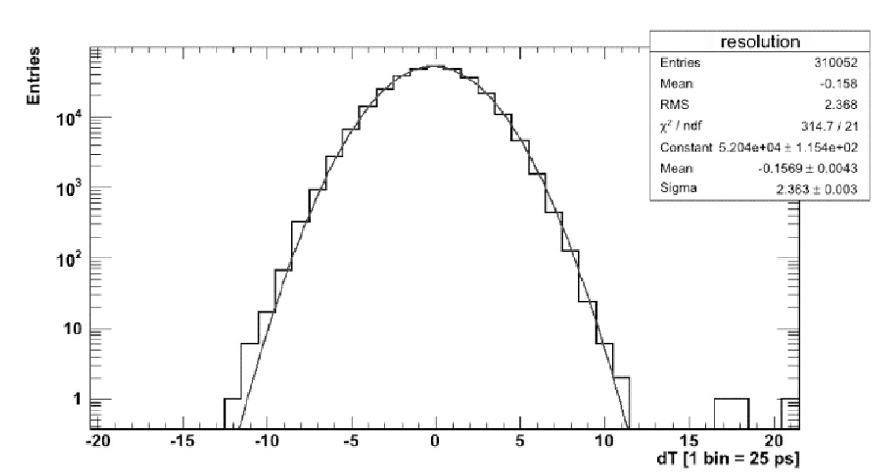}\\
\caption{Distribution of the time difference between contemporary pulses sent along a reference channel and each available channel of electronics, after the offset subtraction.}
\label{fig:Distribution_map}
\end{figure}

In ideal conditions the distribution of the time difference $dT$  between two contemporary pulses in two different channels of the same TDC should correspond to a narrow peak centered at zero.\\
However the dT distribution of all the $64$ available channels with respect to a reference one (Figure \ref{fig:Distribution_no_map}) ranged over $80$ bins (about $2$ ns) due to the electronics intrinsic delay: this value overcame the expected time resolution of each MRPC device of more than one order of magnitude \cite{bib24}.\\
Therefore when a charge is induced on one or more pads, the pulses along each transmission line undertook different systematic delays before being recorded from each TDC module. They mainly depend on the channel, on the setup conditions and on the architecture of TDC chips.\\
The constant systematic uncertainty can be removed by just subtracting an offset $\Delta t_{i}$ for each $t_{i}$.

\begin{equation}
\Delta t_{i} = t_{i} - t_{ref}
\label{eqn:1}
\end{equation}

being $t_{i}$  and $t_{ref}$ the time stamp recorded for a contemporary pulse sent along two different channels. \\
This offset $\Delta t_{i}$ , indeed, corresponds to the mean value of the Gaussian fit of the $dT$ distribution. A single reference channel was chosen for the two TDCs.\\
The contribution to the time resolution due to the electronics setup is calculated by fitting the $dT$ distribution (after offset subtraction) of all the channels connected to each TDC module (Figure \ref{fig:Distribution_map}). Assuming a Gaussian shape, the standard deviation of the distribution has been measured to be $59.5$ ps.\\

\begin{figure}[htpb]
\centering
\subfigure[]{%
\includegraphics[scale=0.9]{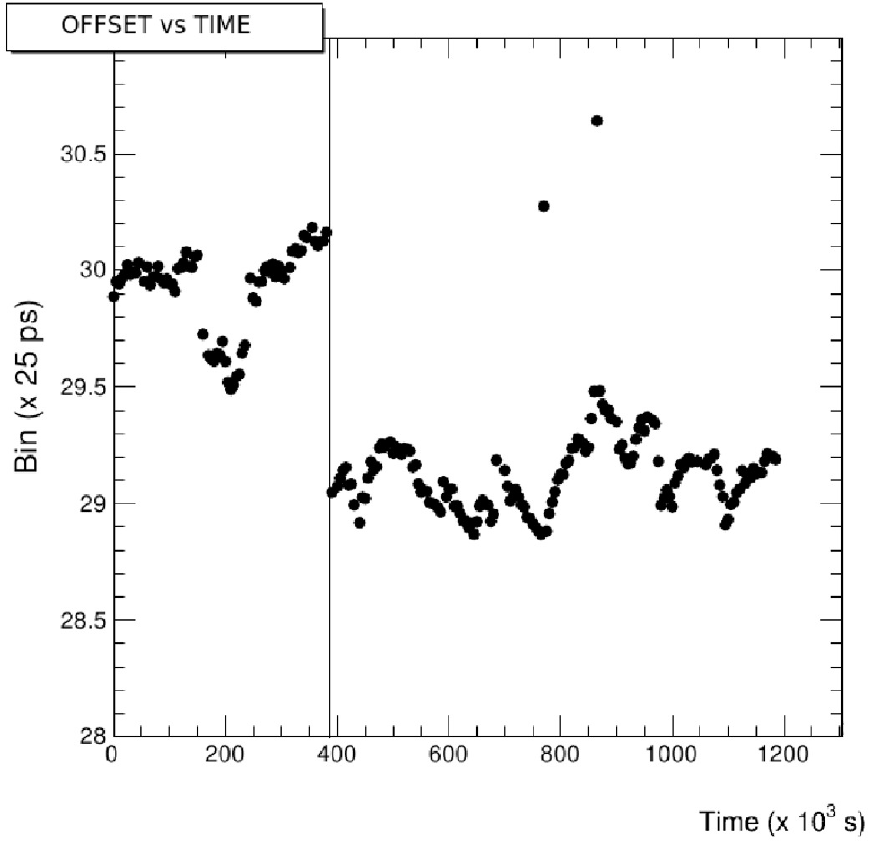}
\label{subfig:6a}}
\qquad
\subfigure[]{%
\includegraphics[scale=0.9]{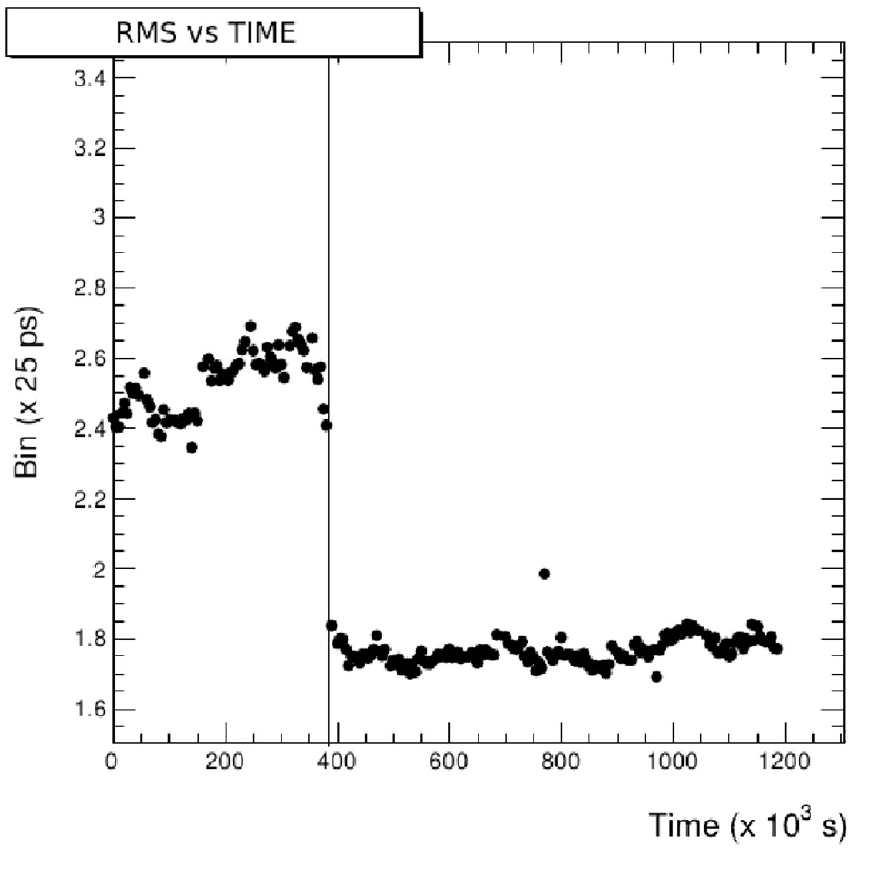}
\label{subfig:6b}}
\caption{a) Example of measured offset of dT distribution for two contemporary pulses, referred to an arbitrary channel pair as a function of time. b) Measured RMS of dT distribution for two contemporary pulses, referred to an arbitrary channel pair, as a function of time. The fluctuations correspond to about $10\%$ of the central value ($5$ ps).}
\label{fig:Stability}
\end{figure}

A study of the stability of the delay values was performed sending a $2$ Hz pulse along couples of selected channels for a period of about $2$ weeks. The test was performed in two steps, separated by a power cycle of the electronics. Both the average value of the delay and the RMS of the channels pairs were acquired and are shown in Figure \ref{subfig:6a} and Figure \ref{subfig:6b} respectively. It is clearly evident the effect of switching off and on again the electronics between the two steps.\\
Nevertheless, before and after the power cycle, the time fluctuations of the mean value of the offset and the RMS were relatively small (about \mbox{$17.5$ ps} and \mbox{$14.5$ ps} respectively) except for slight deviations which are however less than the measured MRPC time resolution.\\

\section{Experimental Setup}\label{SubSec:4}

\begin{figure}[h]
\centering
\includegraphics[scale=1]{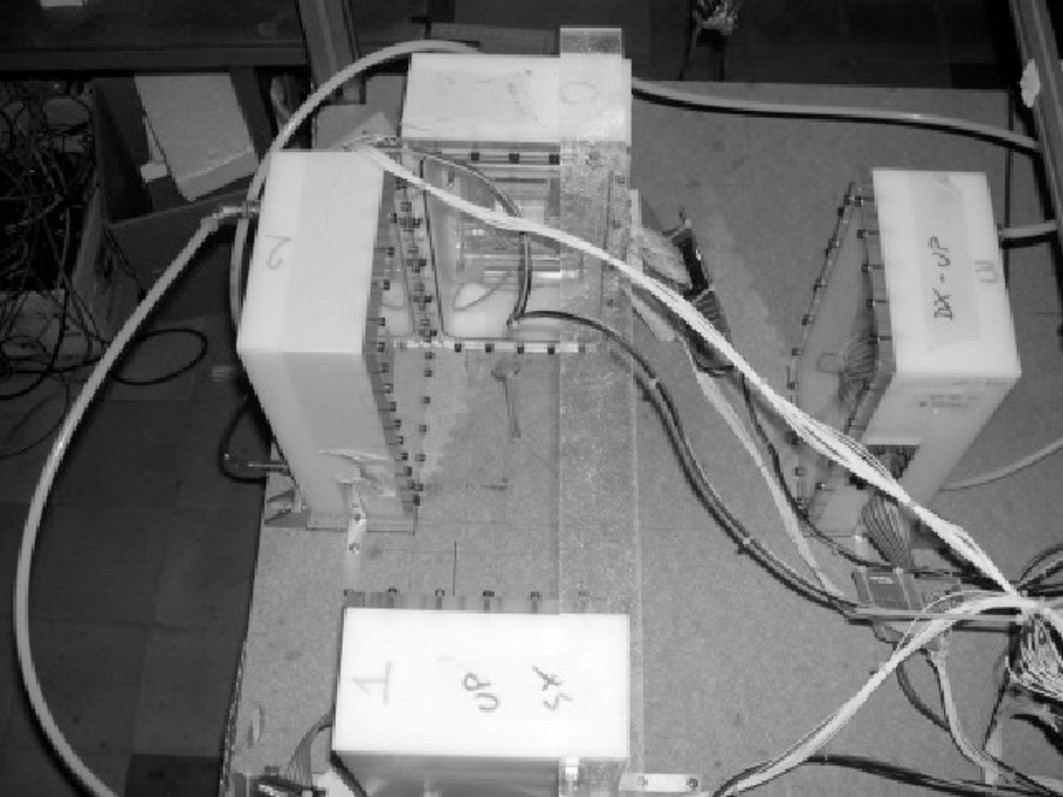}\\
\caption{Experimental setup for the Time of Flight measurement. The $^{22}Na$ source is placed between the two couples of MRPCs detectors. The detectors were placed asymmetrically to introduce a chronological order among the recorded hits.}
\label{fig:Setup}
\end{figure}

Each MRPC detectors pair has been placed at a mutual distance d = $35$ cm, according to an asymmetric cross shape as shown in Figure \ref{fig:Setup}. A $^{22}Na$ gamma source (about $2.5$ $\mu$Ci activity) was placed $9$ cm far from the midpoint of the axis of each detector pair, at the same height of the central pads. \\

It should be noted that the $^{22}Na$ decay takes place with the emission of  two annihilation gammas ($90\%$ BR) and one more energetic ($1.274$ MeV) photon ($100\%$ BR) which are indistinguishable due to the lack of an intrinsic energy resolution for a MRPC.  As a consequence an irreducible background is expected. \\
A photon pair must be detected "in coincidence" on two faced detectors to be considered valid. In this case and ideal Line of Response (LoR) between the two hit points can be created. A "coincidence" is defined by the OR of the two AND signals produced by the two pairs. The trigger was sent to a CAEN TDC V$1290$A front panel and a time window gate armed: all the pulses coming from the readout pads were sent to the TDC buffer through 16 channels flat cables.\\
A hit occurring within the time window opened by the trigger was acquired and the time stamp and the corresponding channel recorded. An event was so defined by the collection of all the hits coming from the four detectors within the same trigger.

\section{Results}\label{Sec:5}

\begin{figure}[htpb]
\centering
\includegraphics[scale=1]{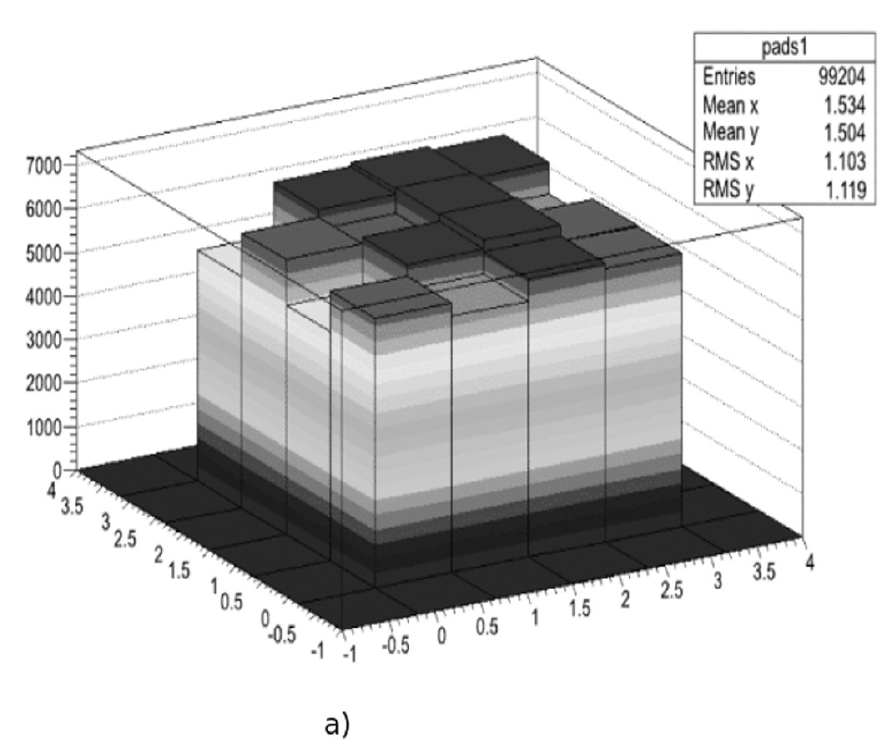}%
\qquad\qquad
\includegraphics[scale=1]{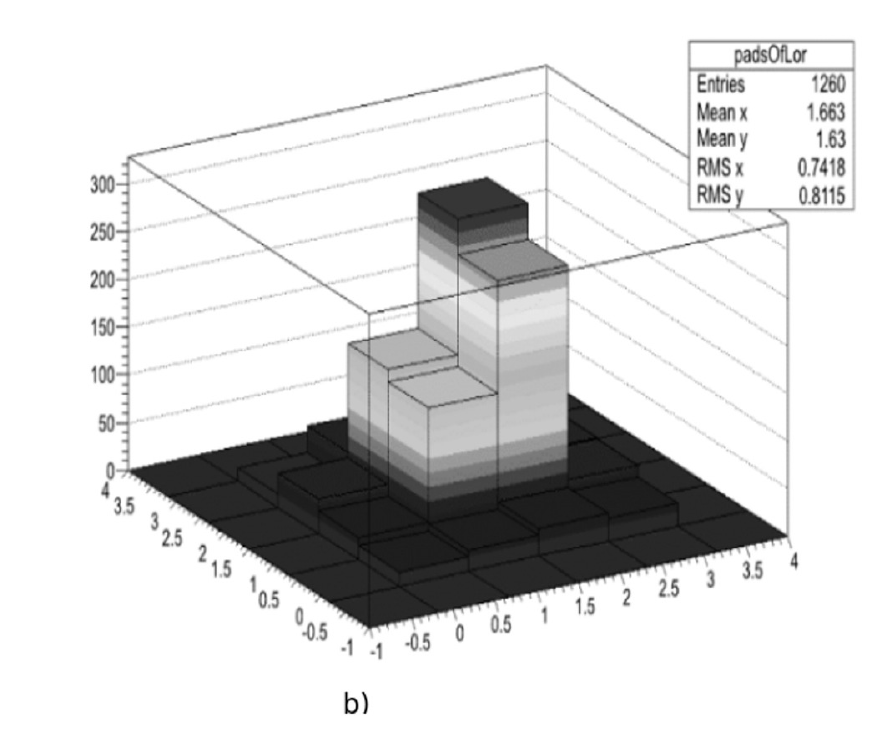}
\caption{2D Occupancy of one of the MRPCS closer to the source a) before and b) after the LOR reconstruction.}
\label{fig:Occupancy}
\end{figure}

Analyzing the occupancy of the detectors closer to the source, before (Figure \ref{fig:Occupancy}a) and after (Figure \ref{fig:Occupancy}b) LoRs reconstruction, it is evident how the maximum activity was found in correspondence of the central pads, i.e. in correspondence of the source position.\\

\begin{figure}[htpb]
\centering
\includegraphics[scale=1]{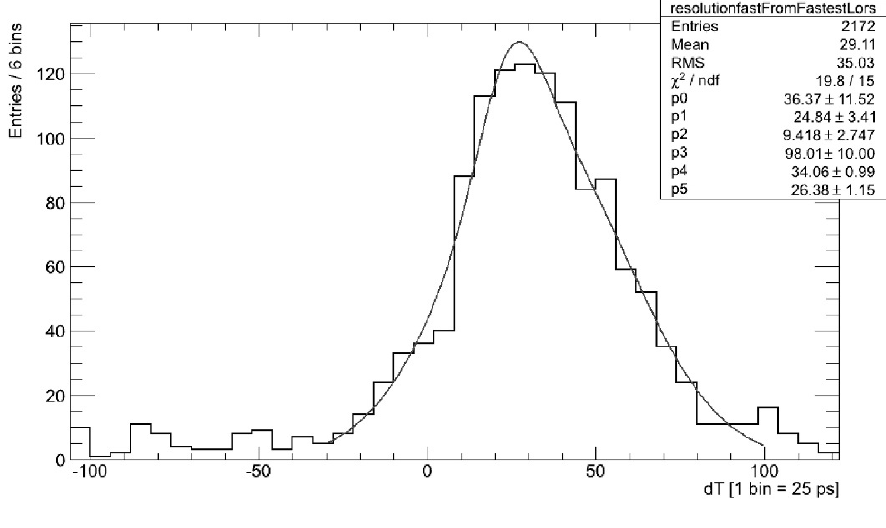}\\
\caption{Time distribution of the fastest LORs for four MRPC detectors. A fit with 2 Gaussians (one due to the direct signal and the other one to the backscattering from the faced detector) has been superimposed.}
\label{fig:Resolution4}
\end{figure}

A constraint on the maximum number of allowed hits per chamber has been introduced to reject other noise sources which could induce multiple spurious signals on clusters of pads. All the events with a number of hits greater than 6 have been rejected.\\
Due to the low efficiency of the device, a tight selection of "good" events has been required to minimize both the contributions of scatter fraction and electronic noise.\\
We have already mentioned that the point source has been shifted from the midpoint of each detector's pair axis. Therefore, for each good event the first hit meeting all the selection requirements, is expected to appear on the detector closest to the point source. A chronological order among recorded hits has been so introduced.\\ 
The time differences $dT$ between the first good hit ($t_{Ch1}$) on one detector and all the subsequent hits recorded ($t_{Ch2}$) on the opposite detector have been collected. The histogram of Figure \ref{fig:Resolution4} has been filled with the least time difference $dT$ of the whole set (in the following referred to as "the faster Lor").\\

Since the time difference $dT$ has been calculated as 

\begin{equation}
dT = t_{Ch2} -t_{Ch1}
\end{equation}

for a point source close to detector $1$, a peak on positive abscissas is expected.
Being the actual source position well known, a ToF-based measurement of the source position was obtained by fitting the signal peak, the sigma of the peak representing the time resolution on the gamma pairs.
According to the source and detectors relative coordinates, the expected time difference (in $25$ ps bins) of two collinear photons along the detector's axis is approximately:
	
\begin{equation}
\Delta t = t_{2} - t_{1} = \frac{0.26m-0.09m}{3\cdot10_{8}m/s} \approx 556 ps \approx 22.6 bins
\label{eqn:deltaT}
\end{equation}
			           
This value does not take into account the relative error with which the detectors are positioned inside the boxes (that we assume to be $2$ mm, i.e. \mbox{$0.3$ bins}). The value of Equation  \ref{eqn:deltaT} is to be considered the true value at which the source has to be reconstructed.  Any reconstructed deviation from this value can be assumed to be the “accuracy” with which the measurement was taken. \\
As mentioned before, the source reconstruction was evaluated from the mean value of the  $dT$ distribution, its peak being dipped in a background generated by photons diffusion. \\
We assumed a Gaussian model for both signal and background and performed a fit whose results are shown in Figure 9. The superimposed fit shows for the signal a mean value of $24.8 \pm 3.0$ bins with a standard deviation of \mbox{$9.4 \pm 2.7 $ bins} while the corresponding values for the background are \mbox{$34.1 \pm 1.0$ bins} and \mbox{$ 26.4 \pm 1.2$  bins}.\\
The errors on fit parameters are just statistical, and they take into account parameters correlations \cite{bib25}.
It should be noted here that, in principle, the contribution to the dT distribution coming from the two detectors pairs should be independent and centered (by construction of the experimental setup) on the same central value. The standard deviation value, hence, should not be influenced. Minimal differences in the setup and in the detectors' response however may be expected leading to a broadening of the distribution. In fact, if we consider the dT distribution coming from only one couple of detectors we obtained a mean value of \mbox{$23.9 \pm 2.5$ bins} with a standard deviation of \mbox{$8.0 \pm 2.3$ bins}.\\
The standard deviation $\sigma_{Sig}$  of the signal, which corresponds to the observed time resolution of the system, contains the contribution from the MRPCs ($\sigma_{MRPC}$), the electronics and channel synchronization ($\sigma_{El} = 59.5$ ps) and from the different path's length of the gamma pairs leaving the source under different angles with respect to the detector's axes ($\sigma_{ToF} = 15$ ps):

\begin{equation}
\sigma^{2}_{Sig}=2\sigma^{2}_{MRPC} + \sigma^{2}_{EL} + 2\sigma^{2}_{ToF}
\label{eqn:eq4}
\end{equation}

\begin{figure}[htpb]
\centering
\includegraphics[scale=1]{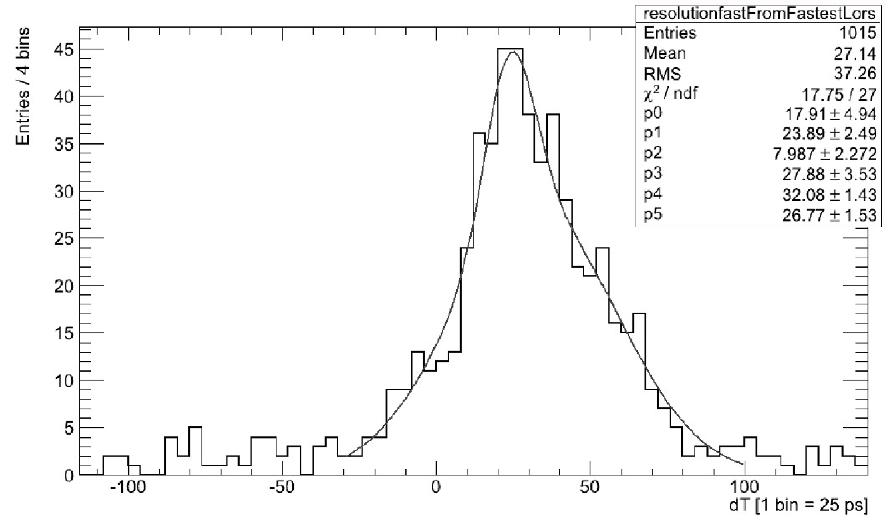}\\
\caption{Time distribution of the fastest LORs for two MRPC detectors. A fit with 2 Gaussians (one due to the direct signal and the other one to the backscattering from the faced detector) has been superimposed.}
\label{fig:Resolution2}
\end{figure}

Subtracting these contributions to the observed time resolution, for a single MRPC we obtain $\sigma_{MRPC} = 160$ ps, corresponding to a FWHM of $376$ ps.
The FWHM is broadened by both the already discussed detector position error inside the boxes ($0.3$ bins = $7.5$ ps) and the TDCs' resolution. This last contribution, which has been measured with a 10 ps resolution pulse generator, amounted to $1.6$ bins ($40$ ps).
By considering only the contribution from one detectors' pair (Figure 10) the obtained value is $\sigma_{MRPC} = 133$ (FWHM = 312 ps), in agreement with the value reported in \cite{bib26} (FWHM $\sim 300$ ps).

\section{Simulation}\label{Sec:6}

\begin{figure}[htpb]
\centering
\includegraphics[scale=1]{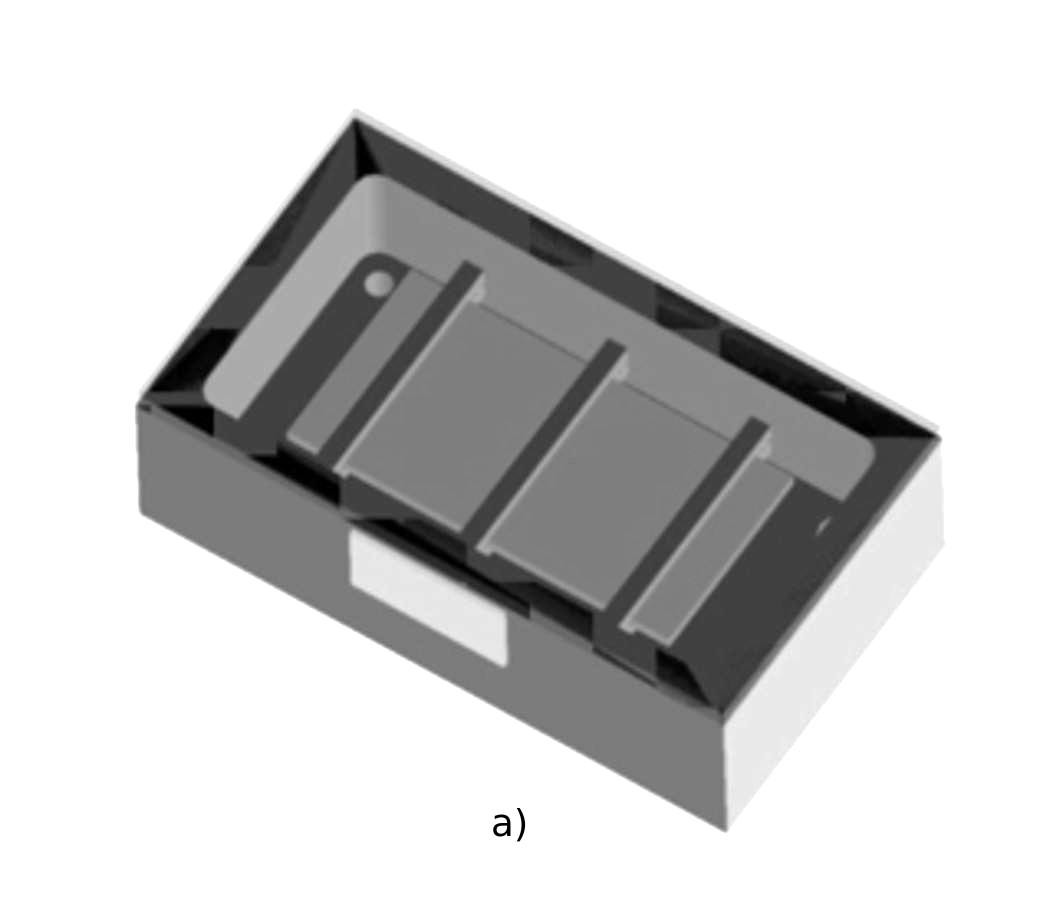}%
\qquad\qquad
\includegraphics[scale=1]{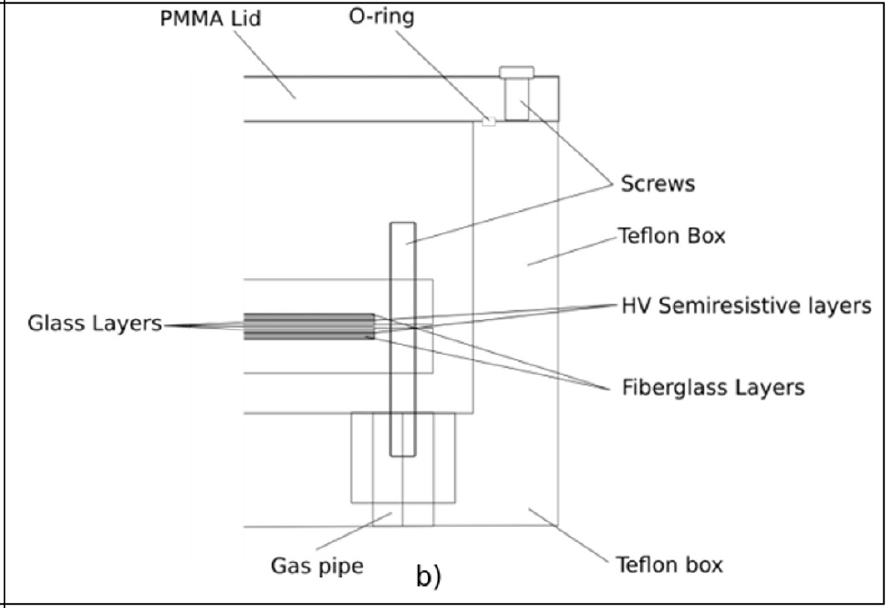}\\
\caption{a) Model of the simulated box and b) detailed representation of the detector scheme.}
\label{fig:SimulatedBox}
\end{figure}

The experimental setup used for the real measurements has been simulated \cite{bib24} through the Geant4 package \cite{bib27}. The number of glass electrodes, the detector position, the surrounding materials, the kind of source, have been reproduced (Figure \ref{fig:SimulatedBox}). Both gammas propagation and electron scattering inside each detector have been studied. The virtual model in Geant$4$ was built assembling simple volumes (Constructing solid Geometry), since more complex techniques (Boundary Represented Solid) made the simulation execution much slower, with a very little improvement of the geometrical accuracy.

The simulated setup was composed by two symmetric MRPCs at distance $d$. A fixed point-like gamma source placed $9$ cm far from the nearest detector, reproducing the exact position of the source in the real experimental setup has been added to.
The generated photons were forced to be emitted back-to-back; nonetheless positron diffusion and annihilation have been neglected. A further \mbox{$1.274$ MeV} gamma photon has been added to take into account also undesired $^{22}Na$ gamma emissions.
Screws, spacers, PMMA lock bars and plastic junctions have not been reproduced in the full-fledge simulation. After few tests it has become evident that their contribution was totally negligible.\\
In fact it is not trivial to precisely model “a priori” the whole experiment, from gamma production to signal induction. Many nice simulations concerning MRPCs physics indeed exist  \cite{bib28, bib29}, but a number of parameters should be tuned in order to produced realistic results. \\
Therefore since we were mainly interested to understand the effects of the experimental setup on the measured values, we made some simplifying assumptions to describe detector physics: a Gaussian response function has been assumed to model both the electronic and the physical detector response; any asymmetric tails in MRPC's time response function have been neglected (this approximation indeed holds since the time resolution of the electronics is of the same order of magnitude of the expected detector's one); representative but quite realistic values for the Gaussian standard deviations  \cite{bib30} have been assumed both for the detectors ($100$ ps) and for electronics ($50$ ps).\\
According to the maximum TDC time resolution achievable, we have adopted a 25 ps binning for dT distributions. \\

Let $t^{*}_{i}$  be the ejection time from the glass of an electron coming from gamma conversion; this value has been extracted with the MRPC time resolution $\sigma_{MRPC}$, giving the final time $t_{i}$ 

\begin{equation}
t_{i}= Gauss (t^{*}_{i},\sigma_{MRPC}).
\end{equation}

As described in the previous section the difference between time stamps matching the trigger window is affected by an uncertainty given by the electronic resolution $\sigma_{El}$.

\begin{equation}
dT^{*}_{Meas}= Gauss (dT,\sigma_{El}).
\end{equation}

Moreover, because of the offset corrections applied on each channel during the real measurement, we added a further offset to each time stamp, given by a Gaussian variable $Gauss(0,\sigma_{Offset})$. 
The final $dT$ ($dT_{Meas}$) reads:

\begin{equation}
dT_{Meas}= Gauss (dT,\sigma_{El}) + Gauss(0,\sigma_{Offset}) + Gauss(0,\sigma_{Offset})
\end{equation}

\begin{figure}[htpb]
\centering
\includegraphics[scale=0.85]{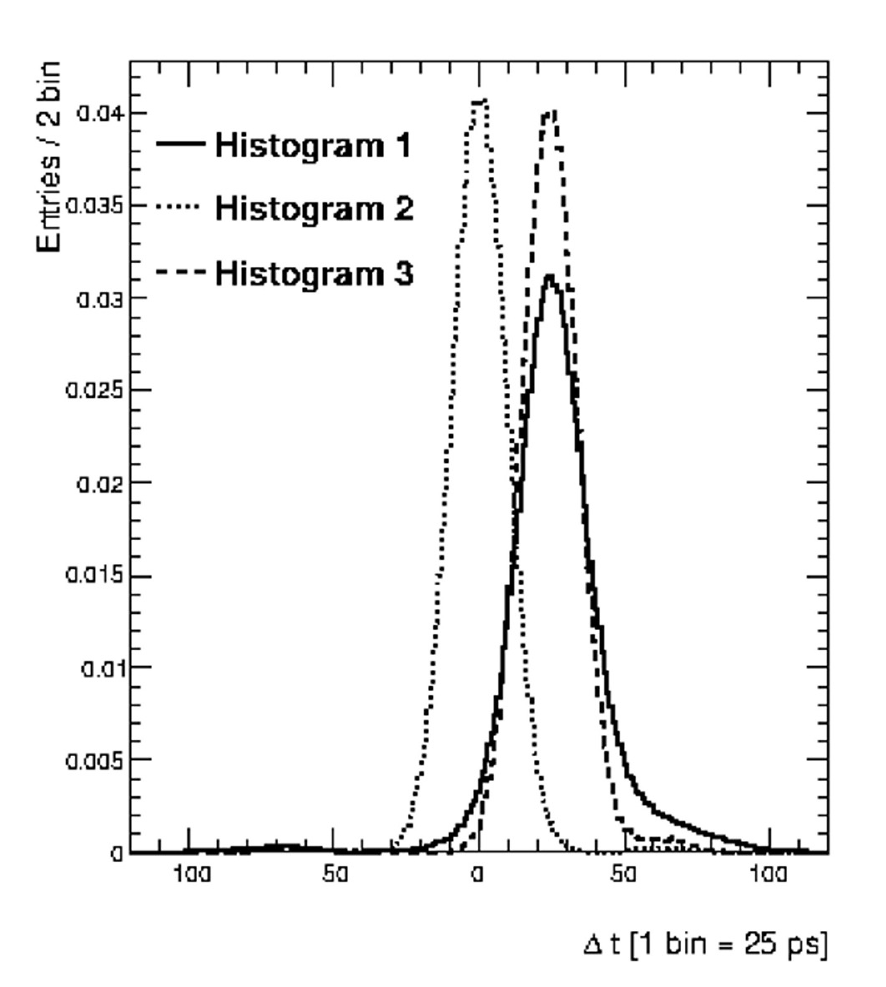}%
\qquad\qquad
\includegraphics[scale=0.85]{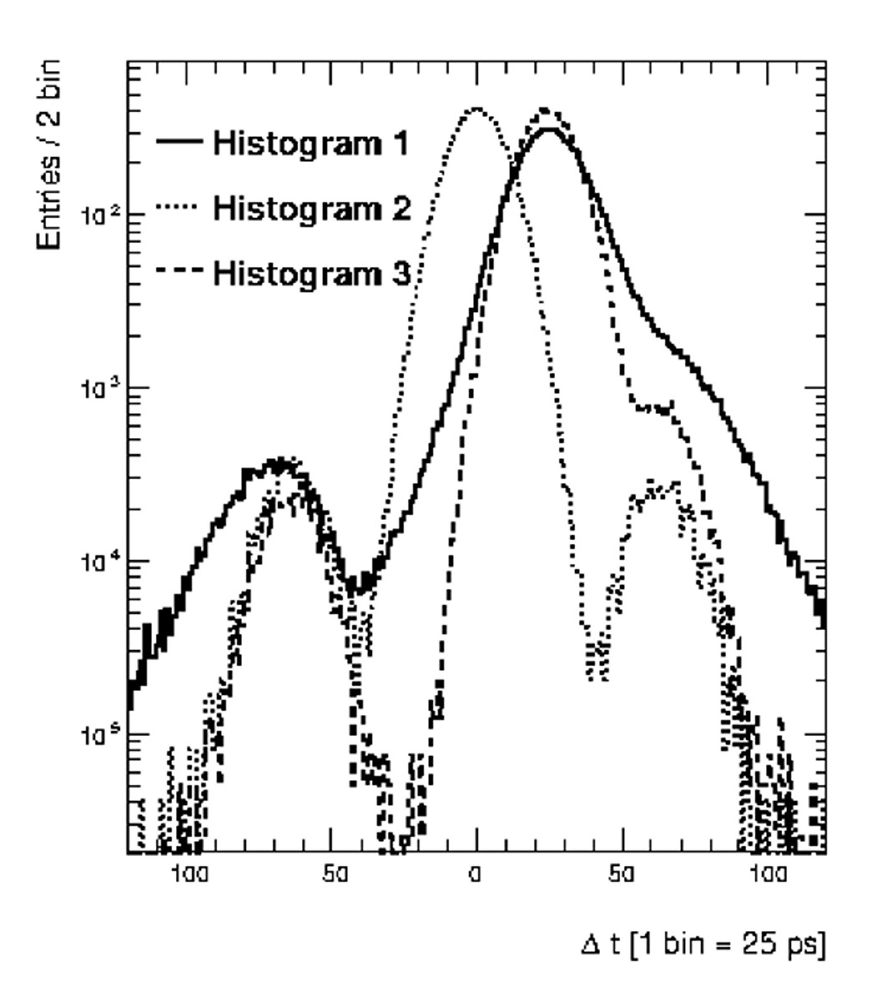}
\caption{$dT$ distributions (in linear and logarithmic scale) obtained by choosing  as the interaction time of photons with glass. Histogram $1$ refers to the complete setup, with Boxes and Lids, while histogram $2$ and histogram $3$ refer to a partial setup, where external boxes have been removed. Two backscattering peaks are visible (in the logarithmic scale) due to gamma scattering against the faced detector case.}
\label{fig:Sim1}
\end{figure}

\begin{figure}[htpb]
\centering
\includegraphics[scale=0.85]{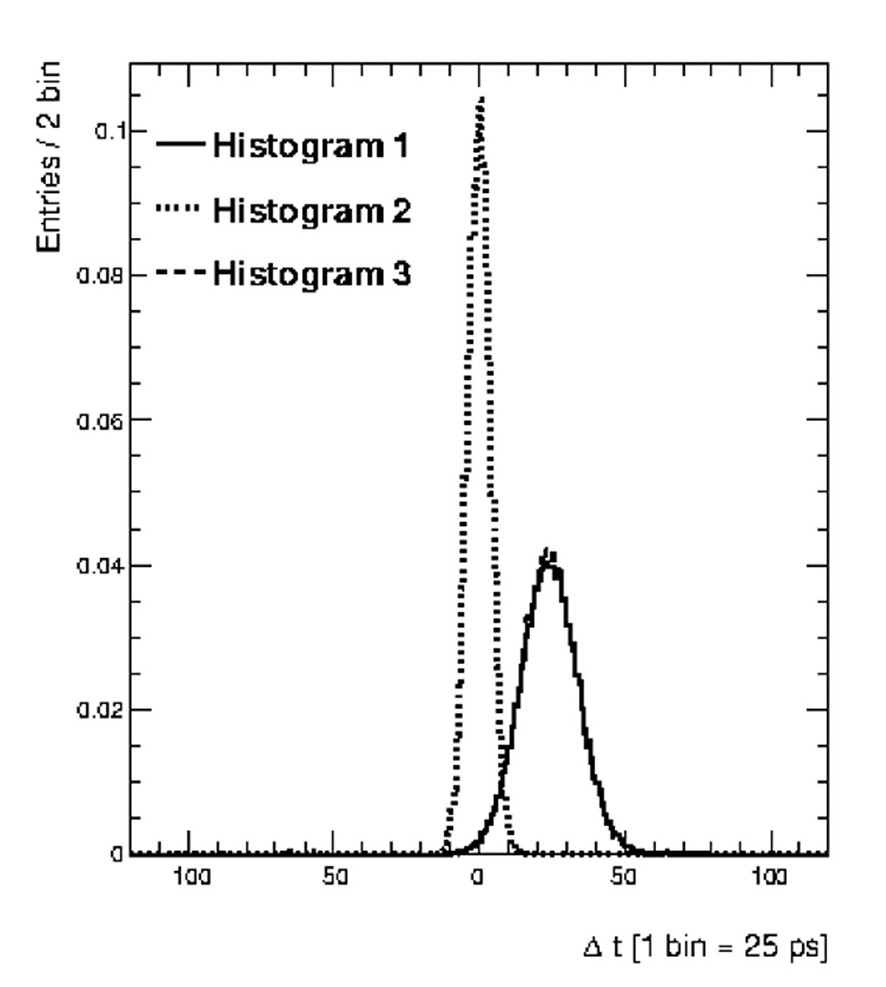}%
\qquad\qquad
\includegraphics[scale=0.85]{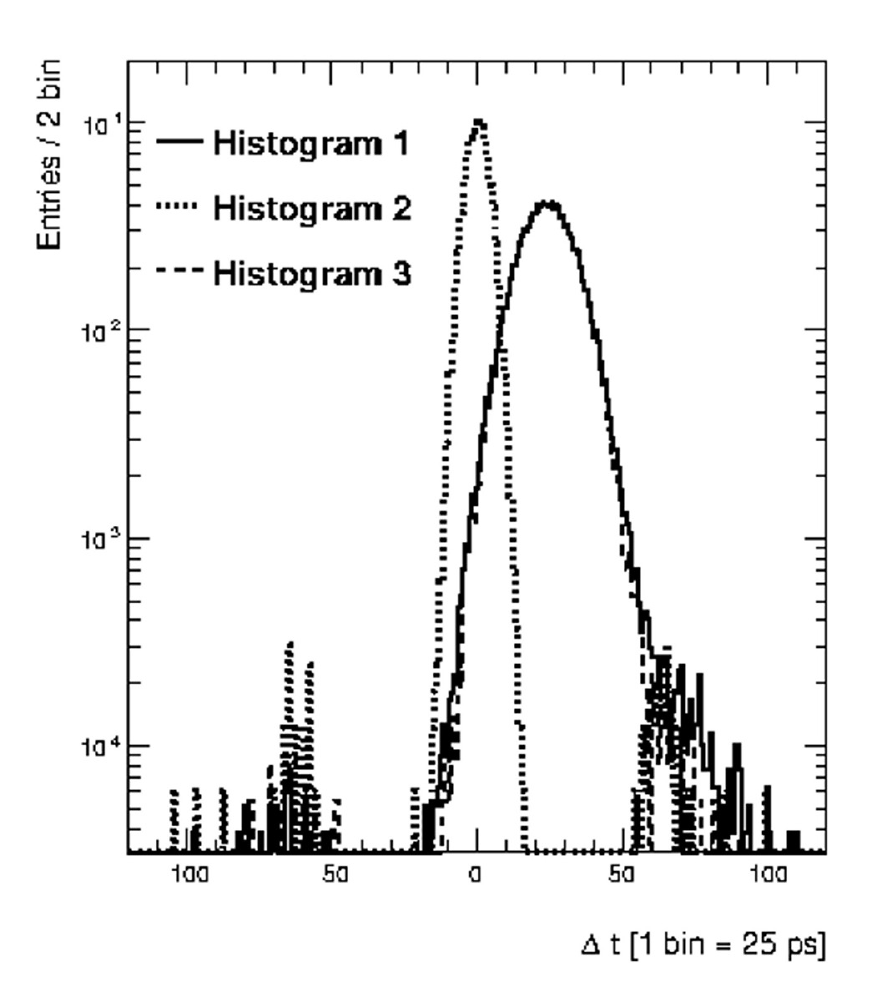}
\caption{$dT$ distributions (in linear and logarithmic scale) obtained by choosing $t_{i}$ as the crossing time of glass-gas interface by single electrons. Histogram $1$ refers to the complete setup, with Boxes and Lids, while histogram $2$ and histogram $3$ refer to a partial setup, where external boxes have been removed.  No evident tails are expected since scattering component is cut by glass sensitivity.}
\label{fig:Sim2}
\end{figure}

\begin{figure}[htpb]
\centering
\includegraphics[scale=1]{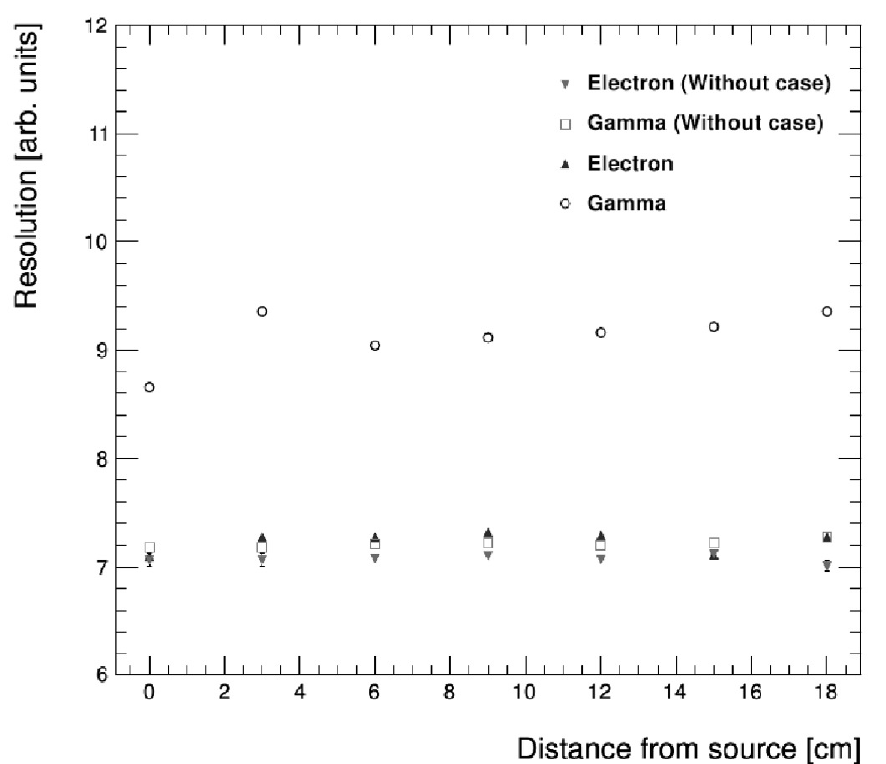}\\
\caption{Trend of resolution as function of distance of the source from the center of the axis. Variations in time resolution are negligible, assuming a detector resolution of about 125 ps.}
\label{fig:Sim3}
\end{figure}

Finally, the standard deviation on $dT$  distribution is calculated by using Equation \ref{eqn:eq4}.
The simulated $dT$  spectra (in linear and logarithmic scale) for two ($4$ gas gaps) MRPCs placed at a distance d = 35 cm, with a point source placed at $9$ cm  away from the center of the axis have been reported in Figure \ref{fig:Sim1} and Figure \ref{fig:Sim2}. The straight lines refer to the complete setup, with Boxes and Lids, while the dashed and dotted lines refer to a partial setup, where Polyethylene boxes have been removed. \\
The distribution of Figure 12 has been produced by choosing $t_{i}$ as the collision time between photon and glass (“Gamma” series) while for the distribution in Figure \ref{fig:Sim3},  corresponds to the ejection time of each electrons from the glass into the gas gap (“Electron” series). 

The $dT$ distributions referring to the complete experimental setup have been fitted using two Gaussian functions, leaving all the parameters floating, and constraining the peak position.
In both cases the main peak is very close to the expected position (bin $24$) (the time difference between flight paths is null, on average, just for a centered point source), even if the distribution spreads out on a wide range. \\
Since MRPCs signals are induced by the electrons' drifting in the gas gaps, just Electron series data should be taken into account from a realistic point of view.
However it is noteworthy that while gamma scattering (Figure \ref{fig:Sim1}) on surrounding experimental apparatus should worsen the time resolution (by about $20\%$), the actual measurement given by the electrons in the same conditions is just negligibly affected by gamma diffusion. \\
Actually the scattering due to the setup worsen the expected time resolution, because it introduces a further indistinguishable contribution due to Compton component. Nevertheless, because of the reduced glass sensitivity to low energy scattered photons, this Compton component does not produce any detectable signals.\\
In Figure \ref{fig:Sim3}, the time resolution (in arbitrary units) as function of source position has been investigated. In this case, every contribution from electronics has been neglected, since it has just assumed to be constant.\\
Due to the relatively large contribution of MRPCs to the total time resolution, even the expected geometrical effects due to the source position are in fact negligible. This confirm that no appreciable worsening of the time resolution is expected moving the source $9$ cm far from the detectors in our experimental conditions.\\

\section{Conclusions}
A time resolution (FWHM) of $312$ ps and $376$ ps has been obtained for a single MRPC by considering respectively the contribution from one and two detectors' pairs. Data were taken with a $50$ mV discrimination threshold and a 15 kV high voltage.
A full-fledge simulation of our experimental setup has been implemented to study the performances of our apparatus and the contribution of experimental setup to the time resolution. A representative value of MRPC time resolution has been assumed, while the measured electronic time resolution ($59.5$ ps) has been used for a convolution of gamma time stamps.\\
The simulation shows that the contribution of misalignment of detector source with respect to the midpoint of detector axis is negligible. The same simulation gives slight different results in presence of the external protective plastic case. The Compton scattering contribution worsens the maximum time resolution achievable with such a device. An improvement of experimental setup may be achieved switching to metal external cases. \\
The measurement and simulation results are consistent with data reported in literature \cite{bib26, bib31}.\\
Improving the device's sensitivity remains mandatory and will be the subject of a following paper. An increase of the current MRPC's efficiency (about $\sim 0.15\%$ for gap) is in fact desirable in order to reduce the acquisition time and improve the signal-to-noise ratio. 

\bibliography{bibfile}

\end{document}